# A New Approach to the Synthesis of Conjugated Polymer: Nanocrystal Composites for Heterojunction Optoelectronics

**Andrew Watt\*, Elizabeth Thomsen, Paul Meredith, and Halina Rubinsztein-Dunlop**
*Soft Condensed Matter Physics Group and Centre for Biophotonics and Laser Science, School of Physical Sciences, University of Queensland, Brisbane, Ausralia. Fax: +61 7 3365 1242 Tel: +61 7 3365 1245; E-mail: watt@physics.uq.edu.au*

**We report a simple one pot process for the preparation of lead sulphide (PbS) nanocrystals in the conjugated polymer poly (2-methoxy-5-(2'-ethyl-hexyloxy)-*p*-phenylene vinylene) (MEH-PPV), and we demonstrate electronic coupling between the two components.**

Conjugated polymer-based heterojunction optoelectronic devices have received much attention recently[1]. In these structures, two active materials are blended to enhance either the optical or electronic properties. Examples of this type of system are polymer blends[2], small organic molecule: polymer blends[3] and nanocrystal: conjugated polymer composites[4]. In this paper we report a new and improved synthesis for the latter system. We envisage applications in plastic photovoltaics and other soft optoelectronic devices.

The current techniques for making nanocrystal: conjugated polymer composite materials rely upon synthesizing nanocrystals separately, and then mixing them with the conjugated polymer[5]. This approach has two shortcomings: firstly, a surfactant must be used to control nanocrystal size and shape. Some of the surfactant becomes incorporated into the final nanocrystal and conjugated polymer mix, which inhibits efficient charge transfer. Secondly, the mixing approach requires the use of co-solvents, which can adversely effects nanocrystal solubility and polymer chain orientation. Our research is focused on solving these synthetic problems.

The major advantage of the new method we describe in this paper is that it eliminates the need for an initial surfactant to terminate nanocrystal growth, and also eliminates the need for subsequent transfer to the conjugated polymer. A similar method has been proposed by Milliron *et al.*[6] which utilizes an electroactive surfactant. Although our method does not allow tight control of nanocrystal size distribution, it does allow more intimate contact between nanocrystal and the conjugated polymer backbone, which we believe will enhance electronic coupling between the two components and hence improve charge transfer in the system. It is also a significantly less complicated synthetic route.

Our novel approach uses the conjugated polymer MEH-PPV (poly (2-methoxy-5-(2'-ethyl-hexyloxy)-*p*-phenylene vinylene)) to control the nanocrystal growth and passivate surface states. MEH-PPV has a high hole mobility and low electron mobility[7]. This relative imbalance limits the performance of any optoelectronic device based upon the material. Nanocrystals, by acting as a percolated high mobility pathway for electrons, offsets this imbalance[8]. In photovoltaic applications, it is thought that photoexcited charge separation occurs at the nanocrystal-polymer interface[9]. Hence, the conjugated polymer acts as a colloidal template, and also as the continuous conductive matrix through which photogenerated charges are transferred to the external circuit. We chose lead sulphide (PbS) as the inorganic material because, in the quantum regime, it has a broad band absorption.[10] Additionally, the electrons and holes are equally confined in PbS nanocrystals, [10] and they been shown to exhibit long excited state lifetimes [11].

The nanoncrystal: conjugated polymer composite was prepared as follows: A sulphur precursor solution was made by dissolving 0.08g of sulphur flakes in 10ml of toluene. The mixture was stirred and degassed with argon for 1 hour. In a typical synthesis, 20ml of toluene, 0.01g of MEH-PPV, 5ml of di-methylsulfoxide DMSO and 0.1g of lead acetate were mixed and degassed with argon at 100 ºC for 2 hours in a 25 ml three-neck flask connected to a Liebig condenser. All materials where purchased from Sigma Aldrich and used without further purification. The resultant solution was bright orange in colour with no precipitate or solvent separation. With the solution at 100 ºC, 1ml of the sulphur precursor was injected. The reaction took approximately 15 minutes to reach completion upon which a brown solution resulted. The product was cleaned to remove excess lead or sulphur ions, DMSO and low molecular weight MEH-PPV by adding the minimum amount of anhydrous methanol to cause precipitation of the composite material. The sample was centrifuged and the supernatant removed. The precipitate was then redissolved in the desired solvent (for example toluene or chlorobenzene). Samples could be taken at any stage and the reaction halted by quenching in toluene at ambient temperature. Typically, solutions produced using these conditions contained ~40% by weight nanocrystals.

Transmission electron microscopy (TEM) was carried out using a Tecnai 20 Microscope. Samples where prepared by taking the cleaned product, diluting it and placing a drop on an ultra thin carbon coated copper grid (Ted Pella) with the Formvar removed. To measure photoluminescence (PL) aliquots were taken at 3 minutes and 15 minutes, cleaned and re-disolved as above, and spun cast on to a 25 x 25 mm pre-cleaned microscope slide. Film thicknesses were measured using a Tencor Alpha-Step 500 Surface Profilometer, and PL measurements where obtained using a Spex Fluoromax 3 spectrometer.

TEM was used to gain an understanding of the nanocrystal growth and quality. Figure 1a shows how the composite material dries into a continuous ultra thin film. A 2μm selected area diffraction pattern was obtained on a flat piece of the film, and the resulting diffraction image can be seen in figure 1b. The diffraction corresponds to the lattice parameter and pattern of cubic PbS looking down the [1,1,1] zone axis. Usually samples prepared from colloidal solutions display only circular poly-crystalline electron diffraction patterns. Examining the film at higher magnification (figure 1c) we see it is composed of individual nanocrystals. This is an important result, as it shows that non-aggregated PbS nanocrystals form. The diffraction pattern in figure 1b would tend to indicate a low degree of orientational anisotropy at the ensemble level. These are similar results to those reported by Berman *et al.*[12] who showed that an ordered array of nanocrystals could form in a polymer matrix. Our nanocrystals are polydisperse with an average size of 4nm (±2nm). Figure 1d confirms the high degree of crystallinity.

For the composite material to be useful in optoelectronic applications there must be electronic coupling between nanocrystals and the conjugated polymer matrix. Figure 2 shows that photoluminescence emission diminishes as the reaction proceeds, i.e. as the nanocrystal concentration increases. In line with the interpretations of Greenham *et al.*[13] and Milliron et al[6] this confirms electronic coupling between nanocrystal and conjugated polymer. The reductions in PL emission that we observe are larger than the associated measurement uncertainty. It is also interesting to note that the spectral shapes remain constant at the three concentrations – this is confirmation that re-absorption effects have been successfully accounted for, although it is still possible that changes in the materials dielectric constant may produce similar effects. Another possible mechanism that could quench the PL signal is the formation of excited state complexes with free lead or sulphur ions. In the absorption spectra we see no evidence of free ions in the system, and hence discount the complexing mechanism. Finally, PL lifetime measurements of MEH-PPV emission show that longer lived MEH-PPV excited states are quenched by the nanocrystals, this strongly supports our electronic coupling hypothesis†.

Nanocrystal growth is dependent on reaction temperature, time, polymer chain length and polymer solvation. In standard nanocrystal synthesis, growth control is derived from a combination of electrostatic effects from the surfactant functional groups (eg phosphine), and the steric effects of the long surfactant chain (typically C18 to C24). MEH-PPV has no charged functional groups which could electrostatically control nanocrystal growth. Therefore we believe that growth is probably influenced by steric effects of the long chain MEH-PPV. It is worthy of note that bulk PbS is formed if there is no polymer present in the reaction mixture.

In conclusion we have demonstrated that it is possible to make nanocrystals in a conjugated polymer by a simple single step process without the need for additional surfactants. The nanocrystals self-assemble, are highly crystalline and are electronically coupled to the conjugated polymer. Although this method seems particularly suited to PbS in MEH-PPV, it could potentially be applied to other sorts of nanocrystals, e.g. CdSe, and other conjugated polymers. Further work is underway to understand the complex dynamics of nanocrystal growth using different polymer molecular weights, purifying the polymer to yield a narrower distribution of molecular weights, and using other solvent systems in a bid to control nanocrystal size and dispersity.

The work was funded by the Australian Research Council.

## Notes and references

**Figures and Captions**

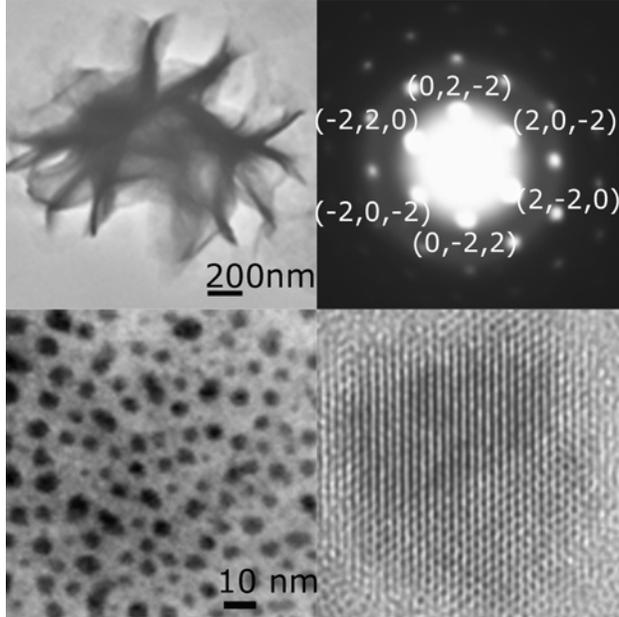

Fig.1 (a) Dark field TEM image of a dried composite film, (b) Selected area (2μm) electron diffraction patern looking down the [1,1,1] zone axis, (c) Scanning TEM image of an ensemble of nanocrystals in a film, (d) Dark field TEM image of a single nanocrystal.

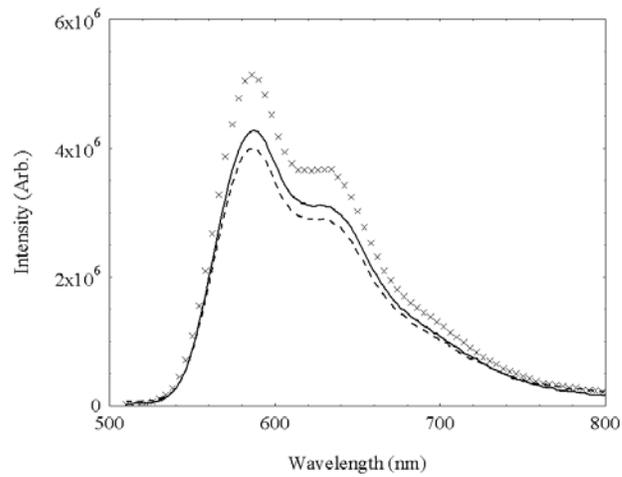

Fig.2 Photoluminescence before injection of sulphur precursor (crosses), 3 min (continuous) and 15 min (dashed). Measurement error ±5.2%.